\documentclass[12pt]{article}
\usepackage[round]{natbib}
\usepackage{amsmath,amssymb,amsfonts} % AA
\usepackage{color}                    % For creating coloured text and background
\usepackage{titletoc}%AA
\usepackage{minitoc,lipsum} %AA
\usepackage{xr}
\usepackage{refcheck}
\usepackage{textcomp, latexsym, parskip,pdflscape,fancyhdr}%subfig,v%AA
\usepackage{subfigure}
\usepackage{mathrsfs}%AA
\usepackage{graphicx}  %\externaldocument[A:]{Introduction.tex}
\usepackage{setspace}
\usepackage{geometry}
\norefnames %AA
\newcommand{\bbeta}{\mbox{\boldmath $\beta$}}

\def\hsigm{\hat{\sigma}}
\def\hsigm2{{\hat \sigma}^2}
\def\Gamma{{\Lambda}}

\def\tbx{ \widetilde{\bf x} }

\def\bY{ {\bf Y} }

\def\bx{ {\pmb x} }
\def\tbx{ \tilde{\pmb x} }

\def\Pr{ {\rm P} }

\def\spacingset#1{\renewcommand{\baselinestretch}%
{#1}\small\normalsize} \spacingset{2}

\def\hbbeta{\hat{\bbeta}}
\def\hG{\hat{G}}

\begin{document}
%\linespread{1.2}
\title{\textbf{Confidence sets for a level set in linear regression }}
\author{
Fang Wan$^{1}$, Wei Liu$^{2}$, Frank Bretz$^{3}$ \\
$^{1}$ Lancaster University, UK\\
$^{2}$ University of Southampton, UK\\
$^{3}$ Novartis Pharma AG, Switzerland\\
}
\date{}
\maketitle

\begin{abstract}
Regression modeling is the workhorse of statistics and there is a vast literature on estimation of the regression function. It is realized in recent years that in regression
analysis the ultimate aim may be the estimation of a level set of the regression function, instead of the estimation of the regression function itself. The published work on estimation
of the level set has thus far focused mainly on nonparametric regression, especially on point estimation. 
In this paper, the construction of confidence sets for the level set of linear regression is considered. In particular, 
$1-\alpha$ level upper, lower and two-sided confidence sets are constructed
for the normal-error linear regression.
It is shown that these confidence sets can be easily constructed from 
the corresponding $1-\alpha$ level simultaneous confidence bands. 
It is also pointed out that the construction method is readily applicable to other parametric regression models where the mean response depends on a linear predictor through a
monotonic link function, which include generalized linear models, linear mixed models
and generalized linear mixed models. Therefore the method proposed in this paper is widely applicable.
Examples are given to illustrate the method. \\

\noindent{\bf keywords}                             % does not recognise keyword ...... is there some other command to put in keywords in my document
 Confidence sets; linear regression; nonparametric regression; parametric regression; simultaneous confidence bands; statistical inference.
\end{abstract}

\vfill
%\hfill {\tiny technometrics tex template (do not remove)}
%%%%%%%%%%%%%%%%%%%%%%%%%%%%%%%%%%%%%%%%%
\newpage
\spacingset{2} % DON'T change the spacing!
\section{INTRODUCTION}
%%%%%%%%%%%%%%%%%%%%%%%%%%%%%%%%%%%%%%%%%
Let $Y = h(\bx) + e$ where $Y \in \Re^1$ is the response, $\bx\in \Re^p$ is the covariate (vector), $h$ is the regression function, 
and $e$ is the random error. 
In regression analysis, there is a vast literature on how to estimate the regression function $h$, based on the observed
data $(Y_i, \bx_i), i=1, \ldots, n$. In recent years, it is realized that an important
problem in regression is the inference of the $\lambda$-level set
$$
G = G_h (\lambda ) = \{ \bx \in K: \, h(\bx) \ge \lambda \}
$$
where $\lambda$ is a pre-specified number, and $K \subset \Re^p$ is a given covariate $\bx$ region of interest. 
It is argued forcefully in Scott and Davenport (2007) that ``In a wide range of regression problems, if it is worthwhile
to estimate the regression function $h$, it is also worthwhile to estimate certain level sets. Moreover, these level sets may
be of ultimate importance. And in many classification problems, labels are obtained by thresholding a continuous variable. Thus,
estimating regression level sets may be a more appropriate framework for addressing many problems that are currently envisioned in other ways".
For example, when considering a regression model of infant systolic blood pressure on birth weight and age, it is of interest to
identify the covariate region over which the systolic blood pressure exceeds (or falls below) a 
pre-specified level $\lambda$. For a regression model
of perinatal mortality rate on birth weigh, it is interesting to identify the range of birth weight over which the 
perinatal mortality rate exceeds a certain $\lambda$. See more details on these two examples in Section 3. Other
possible applications have been pointed out, for example, in Scott and Davenport (2007) and Dau {\it et al.} (2020). 
Inference of the level set $G$ is an important component of the more general field of subgroup analysis (cf. Wang et al., 2007, Herrera et al., 2011, Ting et al., 2020).  
     
In nonparametric regression where $h$ is not assumed to have a specific form, {\bf point} estimation of $G$  aims to construct $\hat{G}$ to approximate $G$ using the observed data.
This has been considered by Cavalier (1997), Polonik and Wang (2005), Willett and Nowak (2007), Scott and Davenport (2007), Dau {\it et al.} (2020) and 
Reeve {\it et al.} (2021) among others. The main focus of these works is on large sample properties such as consistency and rate of convergence. Related work
on estimation of level-sets of a nonparametric density function can be found in Hartigan (1987), Tsybakov (1997), Cadre (2006), Mason and Polonik (2009), Chen {\it et al.} (2017) and Qiao and Polonik (2019).
{\bf Confidence-set} estimation of $G$ aims to construct sets $\hat{G}$ to contain or be contained in $G$ with a pre-specified confidence level $1-\alpha$.  
Large sample approximate $1- \alpha$ confidence-set estimation of $G$ is considered in Mammen and Polonik (2013).

In this paper confidence-set estimation of $G$ for linear regression is considered. It is shown that 
lower, upper and two-sided confidence-set
estimators of $G$ can be easily constructed from the corresponding lower, upper and two-sided  
simultaneous confidence bands for a linear regression function. 
Simultaneous confidence bands for linear regression have been considered in Wynn
and Bloomfield (1971), Naiman (1984, 1986), Piegorsch (1985a,b),  
Sun and Loader (1994), Liu and Hayter (2007) and numerous others; see Liu (2010) for an overview. 
It is also pointed out that the method can be directly extended 
to, for example, the generalized linear regression models, though the confidence-set estimations are 
of asymptotic $1-\alpha$ level since the simultaneous confidence bands are of asymptotic $1-\alpha$ level in this case.
A related problem is the confidence-set estimation of the maximum (or minimum) point of a linear regression model;
see Wan {\it et al.} (2015, 2016) and the references therein.

The layout of the paper is as follows. The construction method of confidence-set estimators is given in Section 2.  The method is illustrated with three examples in Section 3. Section 4 contains conclusions and a brief discussion.
Finally the appendix sketches the proof of a theorem.

\section{Method}

The confidence sets for $G$ are constructed in this section. 
Let the normal-error linear regression model be given by 
$$
Y = h( \bx ) + e = \beta_0 + \beta_1 x_1 + \cdots + \beta_p x_p + e \, ,
$$
where the independent errors $e_i = Y_i - h( \bx_i )$ have distribution $N(0, \sigma^2)$. 
From the observed sample of observations
$(Y_i, \bx_i), i=1, \cdots, n$, the usual estimator of $\bbeta = (\beta_0, \cdots, \beta_p)^T$ is given by 
$\hat{\bbeta} = (X^TX)^{-1} X^T \bY$ where $X$ is the $n \times (p+1)$ design matrix and $\bY = (Y_1, \cdots, Y_n)^T$. The estimator
of the error variance $\sigma^2$ is given by $\hat{\sigma}^2$. It is known that $\hat{\bbeta} \sim N( \bbeta, \sigma^2 (X^TX)^{-1} )$,
$\hat{\sigma}^2 \sim \sigma^2 \chi_{\nu}^2 / \nu$ with $\nu = n-p-1$, and $\hat{\bbeta}$ and $\hat{\sigma}^2$ are independent.
In order for both estimators $\hat{\bbeta}$ and $\hat{\sigma}^2$ to be available, the sample size $n$ must be at least $n \ge p+2$. 

Let $\tbx = (1, \bx^T)^T = (1, x_1, \cdots, x_p)^T$. Suppose the upper, lower 
and two-sided $1-\alpha$ simultaneous confidence bands 
over the covariate region $\bx \in K$ are given, respectively, by  
\begin{eqnarray} 
 \Pr \left \{ \, \tbx^T \bbeta  \le  \tbx^T \hat{\bbeta} + c_1 \hat{\sigma} m( \bx ) \ \forall \, \bx \in K \, \right\} &=& 1- \alpha \,  \\
 \Pr \left \{ \, \tbx^T \bbeta  \ge  \tbx^T \hat{\bbeta} - c_1 \hat{\sigma} m( \bx ) \ \forall \, \bx \in K \, \right\} &=& 1- \alpha \,   \\
 \Pr \left \{ \, \tbx^T \hat{\bbeta} - c_2 \hat{\sigma} m( \bx ) \le \tbx^T \bbeta  \le  \tbx^T \hat{\bbeta} + c_2 \hat{\sigma} m( \bx ) \ \forall \, \bx \in K \, \right\} &=& 1- \alpha \, 
\end{eqnarray}
where $m(\bx ) = \sqrt{ \tbx^T (X^TX)^{-1} \tbx }$ corresponding to the hyperbolic confidence bands, 
and $c_1>0$ and $c_2>0$ are the critical constants to achieve the exact $1-\alpha$ confidence level.
Whilst another popular form is $m( \bx )=1$, corresponding to the constant-width confidence bands, 
the hyperbolic bands are often better than
the constant-width band under various optimality criteria (see, e.g., Liu and Hayter, 2007, and the references therein)
and so used throughout this paper. The critical
constants $c_1$ and $c_2$ can be computed by using the method of Liu {\it et al.} (2005, 2008).

It is worth emphasizing that the three probabilities in (1-3) do not depend on the unknown parameters
$\bbeta \in \Re^{p+1}$ and $\sigma >0$, and that $c_1 < c_2$. 

From the simultaneous confidence bands in (1-3), define the confidence sets as
\begin{eqnarray} 
\hG_{1u} & = & \left \{ \, \bx \in K : \ \tbx^T \hat{\bbeta} + c_1 \hat{\sigma} m( \bx ) \ge \lambda  \, \right\} \, ,   \\
\hG_{1l} & = & \left \{ \, \bx \in K : \ \tbx^T \hat{\bbeta} - c_1 \hat{\sigma} m( \bx ) \ge \lambda  \, \right\} \, ,   \\
\hG_{2u} & = & \left \{ \, \bx \in K : \ \tbx^T \hat{\bbeta} + c_2 \hat{\sigma} m( \bx ) \ge \lambda  \, \right\} ,  \ 
\hG_{2l} = \left \{ \, \bx \in K : \ \tbx^T \hat{\bbeta} - c_2 \hat{\sigma} m( \bx ) \ge \lambda  \, \right\} .  
\end{eqnarray}
The following theorem establishes that $\hG_{1u}$ is an upper, and $\hG_{1l}$ is a lower, confidence set for $G$ of 
exact $1-\alpha$ level, whilst $[\hG_{2l}, \hG_{2u}]$ is a two-sided confidence set for $G$ of at least
$1-\alpha$ level. 
A proof is sketched in the appendix.

\noindent\textbf{Theorem.} We have 
\begin{eqnarray} 
\inf_{\bbeta \in \Re^{p+1},\, \sigma >0} \Pr \left\{ \, G \subseteq \hG_{1u} \, \right\} & = & 1- \alpha \, ,    \\
\inf_{\bbeta \in \Re^{p+1},\, \sigma >0} \Pr \left\{ \, \hG_{1l} \subseteq G \, \right\} & = & 1- \alpha  \, ,  \\
\inf_{\bbeta \in \Re^{p+1},\, \sigma >0} \Pr \left\{ \, \hG_{2l} \subseteq G \subseteq \hG_{2u} \, \right\} & \ge & 1- \alpha  \, .  
\end{eqnarray}

From the definitions in (4-6), it is clear that each set $\hG_{\cdot \cdot}$ is given by all the points in $K$ at which the corresponding simultaneous
confidence band is at least as high as the given threshold $\lambda$. 
Note that each set could be an empty set when $\lambda$ is sufficiently
large, and become $K$ when $\lambda$ is sufficiently small. 
Of course, each set cannot be larger than the given covariate set $K$ from the definition. Since $c_1 >0$ and $c_2 >0$, it is clear that
$\hG_{1l} \subseteq \hG_{1u}$ and $\hG_{2l} \subseteq \hG_{2u}$.  Since $c_1 < c_2 $, it is clear that
$\hG_{1u} \subseteq \hG_{2u}$ and $\hG_{2l} \subseteq \hG_{1l}$. Hence $\hG_{2l} \subseteq \hG_{1l}
\subset \hG_{1u} \subseteq \hG_{2u}$. 

Intuitively, since the regression function $\tbx^T \bbeta$ is bounded from above by the upper simultaneous confidence band $\tbx^T \hat{\bbeta} + c_1 \hat{\sigma} m( \bx )$ over 
the region $\bx \in K$, the level set $G$ cannot be bigger than the set $\hG_{1u}$.  
Similarly, since the regression function $\tbx^T \bbeta$ is bounded from below by the lower simultaneous confidence band $\tbx^T \hat{\bbeta} - c_1 \hat{\sigma} m( \bx )$ over 
the region $\bx \in K$, the level set $G$ cannot be smaller than the set $\hG_{1l}$.
Finally, since the regression function $\tbx^T \bbeta$ is bounded, simultaneously, from below by the lower confidence band $\tbx^T \hat{\bbeta} - c_2 \hat{\sigma} m( \bx )$, and
from above by the upper confidence band $\tbx^T \hat{\bbeta} + c_2 \hat{\sigma} m( \bx )$, over 
the region $\bx \in K$, the level set $G$ must contain the set $\hG_{2l}$ and be contained in the set $\hG_{2u}$ simultaneously. 
%The theorem above asserts that 
%each of the upper, lower and two-sided confidence sets is of confidence level $1-\alpha$.
  
Instead of the level set $G$, the set
\begin{equation}
M = M_h (\lambda ) = \{ \bx \in K: \, h(\bx) \le \lambda \}
\end{equation}
may be of interest in some applications; see e.g. Example 2 in Section 3. 
In this case, one can consider the regression of $-Y$ on $\bx$, given by $-Y = -h( \bx) + (-e)$, and hence $M$ becomes the level set 
$G$ of the regression function $-h(\bx)$
with level $-\lambda$. 

The confidence sets given in (4-6) for the normal-error
linear regression can be generalized to other models that involve a linear predictor 
$\tbx^T\bbeta$. 
In generalized linear models, linear mixed models and generalized linear mixed models (cf. McCulloch and Searle, 2001 and Faraway, 2016), for example, the mean response $E(Y)$ is 
often related to a linear predictor $\tbx^T\bbeta$ by a given monotonic link function $L( \cdot )$, that is, 
$L[E(Y)]=\tbx^T\bbeta$.
Since $L( \cdot )$ is monotone, the set of interest $\{ \, \bx \in K: \ E(Y) \ge L_0 \}$, for a given threshold $L_0$, becomes either $\{ \, \bx \in K: \ \tbx^T\bbeta \ge \lambda \, \}$ or
$\{ \, \bx \in K: \ \tbx^T\bbeta \le \lambda \, \}$, where $\lambda = L( L_0 )$,
depending on whether the function $L( \cdot )$ is increasing or decreasing.
However, when the distribution of $\hbbeta$ is asymptotically normal $N(\bbeta, \hat{\Sigma})$, 
the simultaneous confidence bands of the forms in (1-3) are of approximate $1-\alpha$ level; see, e.g., Liu (2010, Chapter 8).
As a result, the corresponding confidence sets of the forms in (4-6) are of approximate $1-\alpha$ level too. See Example 3
in the next section.

%In logistic regression, for example, the success probability $P=P( \bx)$ is related to
%the covariate $\bx$ by the logit link function $\log \left( P/(1-P) \right) = \tbx^T\bbeta$. Hence the set of interest $\{ \bx \in K: \, P(\bx) \ge P_0 \}$
%becomes $\{ \bx \in K: \, \tbx^T\bbeta \ge \lambda \}$
%of the form $G$ for linear regression, where $P_0$ is a given threshold and $\lambda = \log \left( P_0/(1-P_0) \right)$.  However, only approximate $1-\alpha$
%level simultaneous confidence bands of the forms in (1-3) can be constructed, based on the large sample approximate normal distribution of the estimator $\hbbeta$; see, e.g., Liu (2010, Chapter 8).
%As a result, the confidence sets of the forms in (4-7) are of approximate $1-\alpha$ level too.

\section{Illustrative examples}

In this section, three examples are used to illustrate the confidence sets given in (4-6). 
The \texttt{R} code for all the computation in this section is available from the authors (and will
be made available freely online).

\noindent{\bf Example 1.} In Example 1.1 of Liu (2010), a linear regression model of systolic blood pressure ($Y$) on the two covariates 
birth weight in oz ($x_1$) and age in days ($x_2$) of an infant is considered
$$
 Y = \beta_0 + \beta_1 x_1 + \beta_2 x_2 + e .
$$
Based on the measurements on $n=17$ infants in Liu (2010, Table 1.1), the linear regression model provides a good
fit with $R^2 = 95\%$. The observed values of $x_1$ range from 92 to 149, and the observed values of $x_2$ range from
2 to 5; hence we set 
$K = \{ \, \bx = (x_1, x_2)^T: \ 92 \le x_1 \le 149, 2 \le x_2 \le 5 \, \}$.
It is of interest to identify infants, in terms of
$\bx = (x_1, x_2)^T \in K$, that have mean systolic blood pressure larger than 97, assuming systolic blood pressure larger than 97 is deemed to be too high.
Therefore the level set $G= G(97) = \{ \, \bx \in K: \, \beta_0 + \beta_1 x_1 + \beta_2 x_2 \ge 97 \, \}$
is of interest.

%%%{\bf Fang: please could you write the R code to draw next three 3-d figures and one 2-d figure.}

\begin{figure}[htbp]
\begin{minipage}[t]{0.47\linewidth}
    \centering
    \includegraphics[width=7.5cm]{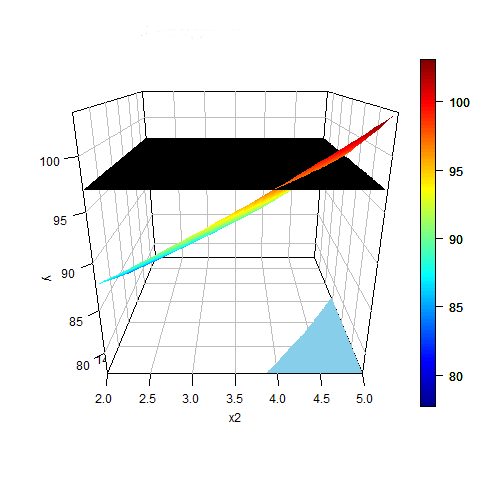}
    \parbox{15.5cm}{\small \hspace{1.2cm}(a) 1-sided upper confidence set $\hG_{1u}$}
\end{minipage}
\hspace{3ex}   
\begin{minipage}[t]{0.45\linewidth}
    \centering
    \includegraphics[width=7.5cm]{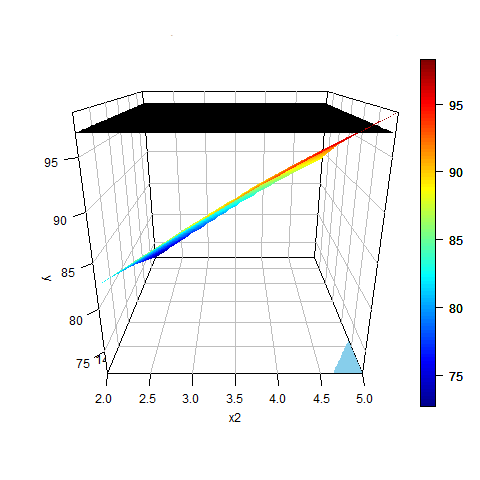}
    \parbox{15.5cm}{\small \hspace{1.2cm}(b) 1-sided lower confidence set $\hG_{1l}$}
\end{minipage}
\vspace{6ex} 
\begin{minipage}[t]{0.47\linewidth}
    \centering
    \includegraphics[width=7.5cm]{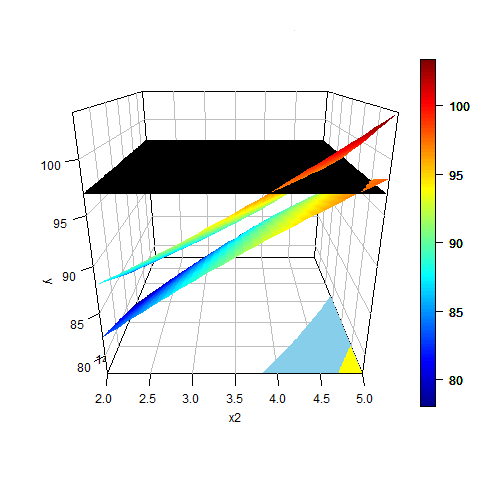}
    \parbox{15.5cm}{\small \hspace{1.2cm}(c) 2-sided confidence set $[\hG_{2l},\hG_{2u}]$}
\end{minipage}
\hspace{3ex} 
\begin{minipage}[t]{0.47\linewidth}
    \centering
    \includegraphics[width=7.5cm]{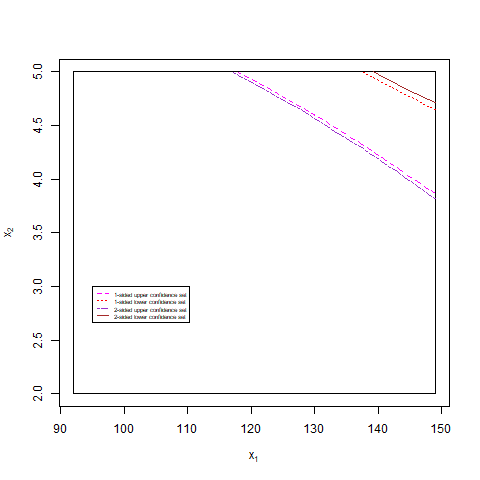}
    \parbox{15.5cm}{\small \hspace{1.2cm}(d) All the confidence sets}
\end{minipage}
%\begin{minipage}[t]{0.45\linewidth}
%\end{minipage}
\begin{center}
\caption{The $95\%$ confidence sets in Example 1, given by the shaded regions. }
\end{center}
\end{figure}

From Section 2, simultaneous confidence bands in (1-3) need to be constructed first in order to construct the confidence sets for $G$ in (4-6).
%The two most popular forms of $m(\bx)$ are $m(\bx ) = \sqrt{ \tbx^T (X^TX)^{-1} \tbx }$ and $m(\bx ) = 1$, 
%corresponding to the hyperbolic and constant-width confidence bands, respectively. As the hyperbolic band is often better than
%the constant-width band under various optimality criteria (see, e.g., Liu and Hayter (2007) and the references therein), hyperbolic
%confidence bands are used in both examples in this section. Now the critical
%constants $c_1$ and $c_2$ can be computed by using the method of Liu {\it et al.} (2005); see also Liu (2010, Section 3.2). 
In this example,
with $p=2$, $n=17$, $1-\alpha = 95\%$ and the given design matrix $X$, $c_2$ is computed 
to be 3.11 and $c_1$ is computed to be $2.77$
by using the method of Liu {\it et al.} (2005) (see also Liu, 2010, Section 3.2).

Figure 1(a) plots the 1-sided
upper confidence set $\hG_{1u}$ in the $\bx$-plane, with the region $K$ given by the rectangle
in solid line. Note that the curvilinear-boundary of $\hG_{1u}$ is given by the projection, to the $\bx$-plane, of the intersection 
between the horizontal plane at height $\lambda = 97$ and the 1-sided
upper simultaneous confidence band over the region $\bx \in K$. The
upper confidence set $\hG_{1u}$ tells us that, with 95\% confidence level, 
only those infants having $\bx \in \hG_{1u}$ may have mean systolic blood pressure larger than or equal to 97. Hence $\bx \in \hG_{1u}$
could be used as a screening criterion for further medical check due to concerns over too high systolic blood pressure.

Similarly, Figure 1(b) plots the 1-sided
lower confidence set $\hG_{1l}$ in the $\bx$-plane. Note that the curvilinear-boundary of $\hG_{1l}$ is given by the projection, to the $\bx$-plane, of the intersection 
between the horizontal plane at height $\lambda = 97$ and the 
1-sided lower simultaneous confidence band over the region $K$. The
lower confidence set $\hG_{1l}$ tells us that, with 95\% confidence level,  
infants having $\bx \in \hG_{1l}$ have mean systolic blood pressure larger than or equal to 97. Hence these infants should 
have further medical check due to concerns over excessive high systolic blood pressure. 

Figure 1(c) plots the two-sided 
confidence set $[\hG_{2l}, \hG_{2u}]$ in the $\bx$-plane. Note that the curvilinear-boundaries 
of $[\hG_{2l}, \hG_{2u}]$ are given by the projection, to the $\bx$-plane, 
of the intersection between the horizontal plane at height $\lambda = 97$ and the two-sided confidence band over the region $K$. The two-sided confidence set tells 
us that, with 95\% confidence level,  infants having $\bx \in K \backslash \hG_{2u}$ are not of concern, infants having $\bx \in \hG_{2l}$ are of concern, and infants having $\bx \in \hG_{2u}$ are
possibly of concern, in terms of
excessive high mean systolic blood pressure.

Figure 1(d) plots $\hG_{1u}$, $\hG_{1l}$ and $[\hG_{2l}, \hG_{2u}]$ in the same picture for the purpose of comparison. 
It is clear from the figure that $\hG_{2l} \subseteq \hG_{1l} \subseteq \hG_{1u} \subseteq \hG_{2u}$
as pointed out in Section 2.

Note that when $\lambda$ is large, 100 say, the horizontal plane at height $\lambda $ and the 
1-sided lower simultaneous confidence band do not intersect over the region $K$.
In this case the 1-sided lower confidence set $\hG_{1l}$ is an empty set. 
Similar observations hold for other confidence sets.

\noindent{\bf Example 2.} Selvin (1998, p224) provided a data set on perinatal mortality
(fetal deaths plus deaths within the first month of life) rate
(PMR) and birth weight (BW) collected in California in 1998. The
interest is on modelling how PMR changes with BW; Selvin (1998)
considered fitting a 4th order polynomial regression model between
$Y=\log(-\log(PMR))$ and $x=BW$: 
$$
Y = \beta_0 + \beta_1 x + \beta_2 x^2  + \beta_3 x^3 + \beta_4 x^4 + e \, .
$$
Here we will focus on the black
infants only, using the 35 observations extracted from Selvin (1998) and
 given in Liu (2010, Table 7.1).
The 4th order polynomial regression model provides a good
fit with $R^2 = 97\%$. 

The observed values of $x$ range from 0.85 to 4.25 and so we set $K = [0.85, 4.25]$.
we are interested in the values of $x \in K$ that may result in excessive high PMR.
Since $Y=\log(-\log(PMR))$ and $\log(-\log( \cdot ))$ is monotone decreasing, we are
interested in the set $M = \{ x \in K: \ \beta_0 + \beta_1 x + \cdots + \beta_4 x^4 \le \lambda \}$ in (10), with
$\lambda = \log(-\log(0.01)) = 1.527$ assuming that $PMR \ge 0.01$ is regarded as excessively high.

\begin{figure}[htbp]
\begin{minipage}[t]{0.48\linewidth}
    \centering
    \includegraphics[width=7.5cm]{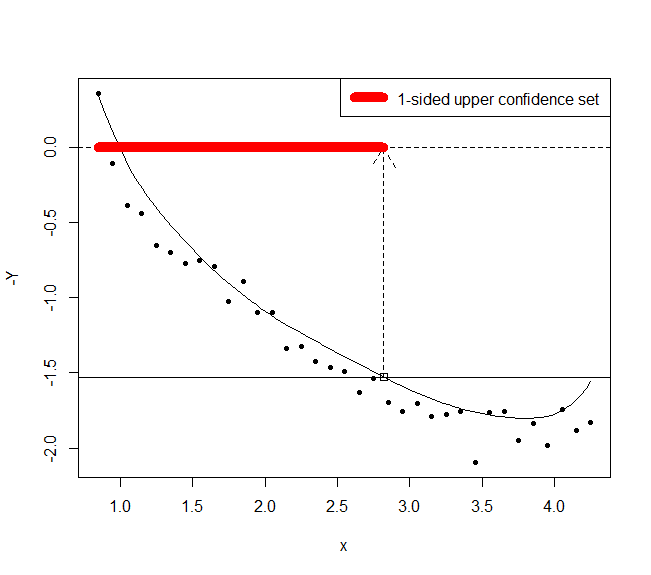}
    \parbox{15.5cm}{\small \hspace{1.2cm}(a)  1-sided upper confidence set $\hG_{1u}$}
\end{minipage}
\hspace{1ex}   
\begin{minipage}[t]{0.48\linewidth}
    \centering
    \includegraphics[width=7.5cm]{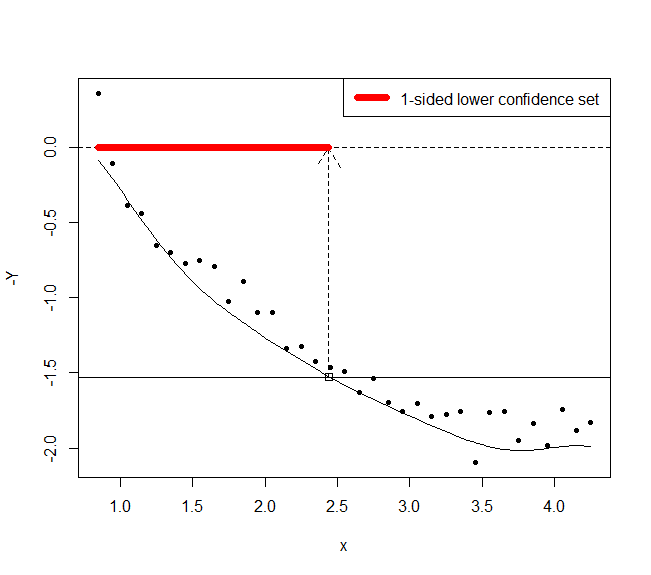}
    \parbox{15.5cm}{\small \hspace{1.2cm}(b)  1-sided lower confidence set $\hG_{1l}$}
\end{minipage}
\vspace{6ex} 
\begin{minipage}[t]{0.48\linewidth}
    \centering
    \includegraphics[width=7.5cm]{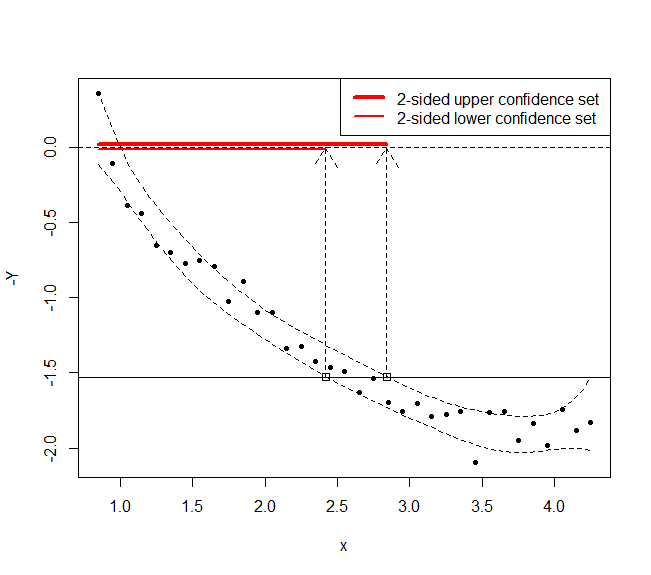}
    \parbox{15.5cm}{\small \hspace{1.2cm}(c)  2-sided confidence set $[\hG_{2l}, \hG_{2u}]$}
\end{minipage}
\hspace{1ex} 
\begin{minipage}[t]{0.48\linewidth}
    \centering
    \includegraphics[width=7.4cm]{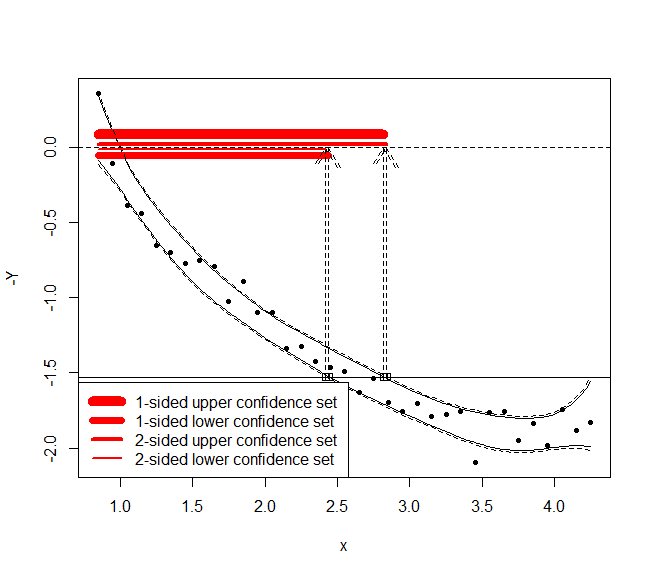}
    \parbox{15.5cm}{\small \hspace{1.2cm} (d) All the confidence sets}
\end{minipage}
\begin{center}
\caption{The $95\%$ confidence sets in Example 2, indicated by the red line segments.}
\end{center}
\end{figure}

From Section 2, simultaneous confidence bands for $-\beta_0 - \beta_1 x - \cdots - \beta_4 x^4$ of the forms 
in (1-3), with $\tbx = (1, x, \cdots, x^4)^T$ in this example, 
over $x \in K$ need to be constructed first in order to construct the confidence sets for $M$ in (10), which is the same as
$G  = \{ x \in K: \, -\beta_0 - \beta_1 x - \cdots -\beta_4 x^4 \ge -\lambda \}$.
The critical
constants $c_1$ and $c_2$   
can be computed by using the method of Liu {\it et al.} (2008); see also Liu (2010, Section 7.1). 
With $p=4$, $n=35$, $1-\alpha = 95\%$ and the given design matrix $X$, $c_2$ is computed to be 2.99 and $c_1$ is computed to be $2.69$. 

Figure 2(a) plots the 1-sided
upper simultaneous confidence band $-\tbx^T \hat{\bbeta} + c_1 \hat{\sigma} m( \bx ) \ {\rm over} \, x \in K$, and the 1-sided
upper confidence set $\hG_{1u} = [0.85, 2.82]$ on the $x$-axis. Note that the boundary of $\hG_{1u}$, 2.82, is given by the projection, to the $x$-axis, of the intersection 
between the horizontal line at height $-\lambda = -1.527$ and the upper simultaneous confidence band over the interval $x \in K$. The
upper confidence set $\hG_{1u}$ tells us that, with confidence level 95\%, only those infants having $x \in \hG_{1u}$ may have $E\{\log(-\log(PMR))\} \le 1.527$ and hence may 
need extra medical care due to concerns over excessive high PMR.

Similarly, Figure 2(b) plots the 1-sided
lower simultaneous confidence band $-\tbx^T \hat{\bbeta} - c_1 \hat{\sigma} m( \bx )$ over
$x \in K$, and the 1-sided
lower confidence set $\hG_{1l} = [0.85, 2.44]$ on the $x$-axis. Note that the boundary of $\hG_{1l}$, 2.44, is given by the projection, to the $x$-axis, of the intersection 
between the horizontal line at height $-\lambda = -1.527$ and the lower simultaneous confidence band over the interval $K$. The
lower confidence set $\hG_{1l}$ tells us that, with confidence level 95\%, infants having $x \in \hG_{1l}$ have $E\{\log(-\log(PMR))\} \le 1.527$ and so should have extra medical care 
due to concerns over excessive high PMR. 

Figure 2(c) plots the 2-sided simultaneous confidence band $[-\tbx^T \hat{\bbeta} - c_2 \hat{\sigma} m( \bx ),\, -\tbx^T \hat{\bbeta} + c_2 \hat{\sigma} m( \bx )]$ over $x \in K$, and the
2-sided confidence set $[\hG_{2l}, \hG_{2u}] = \left[[0.85,2.42], [0.85, 2.84] \right]$ on the $x$-axis. Note that the boundaries of $[\hG_{2l}, \hG_{2u}]$, 2.42 and 2.84,
 are given by the projection, to the $x$-axis, 
of the intersection between the horizontal line at height $-\lambda = -1.527$ and the two-sided confidence band over the interval $K$. The two-sided confidence set tells 
us that, with confidence level at least 95\%, infants having $x \in K \backslash \hG_{2u}$ are not of concern, infants having $x \in \hG_{2l}$ are of concern, and infants having $x \in \hG_{2u}$ are
possibly of concern, in terms of
excessive high PMR.

Figure 2(d) plots $\hG_{1u}$, $\hG_{1l}$ and $[\hG_{2l}, \hG_{2u}]$ in the same picture for the purpose of comparison. 
From this figure, it is clear again that $\hG_{2l} \subseteq \hG_{1l} \subseteq \hG_{1u} \subseteq \hG_{2u}$
as pointed out in Section 2.

\noindent{\bf Example 3.} 
Myers {\it et al.} (2002, p114) provided a data set on
a single quantal bioassay of a toxicity experiment. The
 effect of different doses of nicotine on the
common fruit fly is investigated by fitting a logistic regression model between
the number of flies killed $y$ and $x =\ln ({\rm concentration\ of\ nicotine})$.
Seven
observations of $(y_j, m_j,  x_j)$ are given (see also Liu, 2010, Table 8.1), where
$m_j$ is the number of flies experimented at dose $x_j$. 

Let $p(x)$ denotes
the probability that a fly will be killed at dose $x$. Then $y_j \sim {\rm Binomial} \left( m_j, p(x_j) \right)$
with ${\rm logit} \left( p(x_j) \right) = \beta_0 + \beta_1 x_j$, $j=1, \cdots, 7$.
Based on the seven observations, the MLE $\hat{\bbeta} = (\hat{\beta}_0, \hat{\beta}_1)^T$ is calculated to be $(3.124, 2.128)^T$ and 
the approximate covariance matrix of $\hat{\bbeta}$ is
$$
\hat{\cal I}^{-1} = \left( \begin{array}{cc}
                     0.1122 & 0.0679 \\
                     0.0679 & 0.0490 \\
										       \end{array} \right) \, .
$$
Hence $\hat{\bbeta}$ has approximate normal distribution $N_2 (\bbeta , \hat{\cal I}^{-1})$. 

The seven observed $\left( {\rm logit} (y_j / m_j), x_j \right)$ are plotted
in Figure 3(a), from which it seems that the logistic regression
model fits the observations very well. Indeed the deviance is ${\cal D} = 0.734$, which
is very small in comparison with $\chi_{5, \alpha}^2$ for any conventional
$\alpha$ value. %Hence there is no evidence of lack of fit in the logistic model.

The median effective dose $ED_{50}$ is the dose $x$ such that $p(x) = 0.5$, and often of interest in
dose-response study.  Suppose we are
interested in the doses $x$, within the dose range $K = [-2.30, -0.05]$ of the study, such that $p(x) \ge 0.5$, that is, we want to
identify the level set $G = \{ x \in K: \ \beta_0 + \beta_1 x \ge \lambda \}$ with $\lambda = {\rm logit}(0.5) = 0$ due to
the fact that ${\rm logit} (p)$ is monotone increasing in $p \in (0,1)$.
Now the method of Section 2 can be used to construct various confidence sets for $G$.

\begin{figure}[htbp]
\begin{minipage}[t]{0.48\linewidth}
    \centering
    \includegraphics[width=7.5cm]{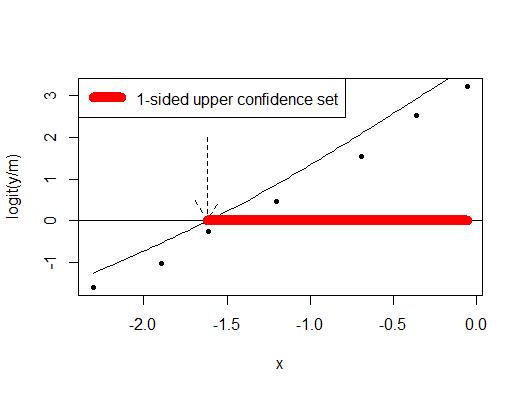}
    \parbox{15.5cm}{\small \hspace{1.2cm}(a)  1-sided upper confidence set $\hG_{1u}$}
\end{minipage}
\hspace{1ex}   
\begin{minipage}[t]{0.48\linewidth}
    \centering
    \includegraphics[width=7.5cm]{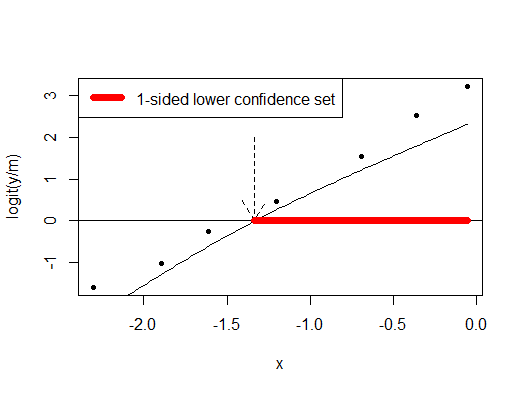}
    \parbox{15.5cm}{\small \hspace{1.2cm}(b)  1-sided lower confidence set $\hG_{1l}$}
\end{minipage}
\vspace{6ex} 
\begin{minipage}[t]{0.48\linewidth}
    \centering
    \includegraphics[width=7.5cm]{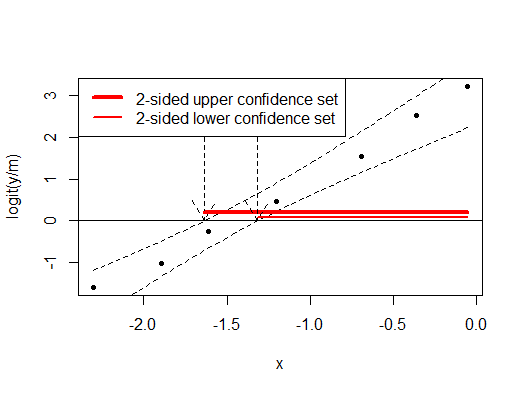}
    \parbox{15.5cm}{\small \hspace{1.2cm}(c)  2-sided confidence set $[\hG_{2l}, \hG_{2u}]$}
\end{minipage}
\hspace{1ex} 
\begin{minipage}[t]{0.48\linewidth}
    \centering
    \includegraphics[width=7.4cm]{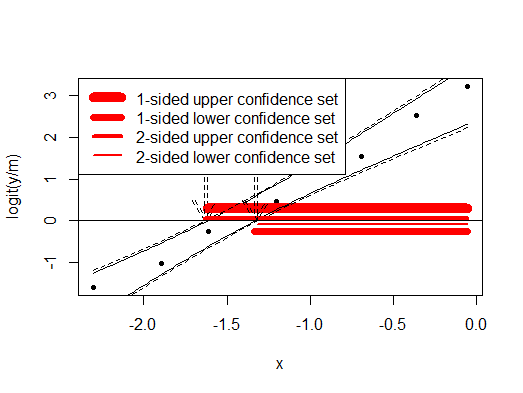}
    \parbox{15.5cm}{\small \hspace{1.2cm} (d) All the confidence sets}
\end{minipage}
\begin{center}
\caption{The $95\%$ approximate confidence sets in Example 3, indicated by the red line segments.}
\end{center}
\end{figure}

From Section 2, simultaneous confidence bands for $\beta_0 + \beta_1 x $ 
over $x \in K$ need to be constructed first in order to construct the confidence sets for $G$.
Note, however, only approximate $1- \alpha$ confidence bands of the forms in (1-3), with $\hat{\sigma} = 1$ and 
$(X^TX)^{-1}$ replaced with $\hat{\cal I}^{-1}$, can be constructed by using the approximate normal distribution
$N_2 (\bbeta , \hat{\cal I}^{-1})$ of $\hat{\bbeta}$. Hence the confidence sets for $G$ are also of approximate 
$1- \alpha$ level.
For $1-\alpha = 95\%$ and $K = [-2.30, -0.05]$, $c_2$ is computed to be 2.42 and $c_1$ is computed to be $2.14$
 by using the method of Liu {\it et al.} (2008); see also Liu (2010, Section 8.2). 

Figure 3(a) plots the approximate 1-sided
upper simultaneous confidence band $\tbx^T \hat{\bbeta} + c_1 \sqrt{ \tbx^T \hat{\cal I}^{-1} \tbx }$ over
$x \in K$, and the approximate 1-sided
upper confidence set $\hG_{1u} = [-1.61, -0.05]$ on the $x$-axis. As before the boundary of $\hG_{1u}$, -1.61, is given by the projection, to the $x$-axis, of the intersection 
between the horizontal line at height $\lambda = 0$ and the upper simultaneous confidence band over the interval $x \in K$. The
upper confidence set $\hG_{1u}$ tells us that, with approximate confidence level 95\%, only those doses $x$ in $\hG_{1u}$ may have 
$p(x) \ge 0.5$. 

Similarly, Figure 3(b) plots the approximate 1-sided
lower simultaneous confidence band $\tbx^T \hat{\bbeta} - c_1 \sqrt{ \tbx^T \hat{\cal I}^{-1} \tbx }$ over
$x \in K$, and the approximate 1-sided
lower confidence set $\hG_{1l} = [-1.33, -0.05]$ on the $x$-axis. 
The lower confidence set $\hG_{1l}$ tells us that, with approximate confidence level 95\%, doses $x$ in $\hG_{1l}$ 
have $p(x) \ge 0.5$. 

Figure 3(c) plots the approximate 2-sided simultaneous confidence band $[\tbx^T \hat{\bbeta} - c_2 \sqrt{ \tbx^T \hat{\cal I}^{-1} \tbx },\, \tbx^T \hat{\bbeta} + c_2 \sqrt{ \tbx^T \hat{\cal I}^{-1} \tbx }]$ over $x \in K$, and the corresponding approximate 
2-sided confidence set $[\hG_{2l}, \hG_{2u}] = \left[[-1.32,-0.05], [-1.64, -0.05] \right]$ on the $x$-axis. The two-sided confidence set tells 
us that, with approximate confidence level at least 95\%, doses $x$ in $\hG_{2u}$ may have, and doses $x$ in $\hG_{2l}$ have, $p(x) \ge 0.5$.

Figure 3(d) plots $\hG_{1u}$, $\hG_{1l}$ and $[\hG_{2l}, \hG_{2u}]$ in the same picture for the purpose of comparison. 
From this figure, it is clear again that $\hG_{2l} \subseteq \hG_{1l} \subseteq \hG_{1u} \subseteq \hG_{2u}$
as pointed out in Section 2, though the differences between $\hG_{2l}$ and $\hG_{1l}$, and between $\hG_{1u}$ and $\hG_{2u}$, are
very small.

%%%%%%%%%%%%%%%%%%%%%%%
\section{CONCLUSION AND DISCUSSION}
%%%%%%%%%%%%%%%%%%%%%%
In this paper, the construction of confidence sets for the level set of linear regression is discussed. Upper, lower and two-sided confidence sets of level $1-\alpha$ are constructed for the
normal-error linear regression.
It is shown that these confidence sets are constructed from the corresponding $1-\alpha$ level simultaneous confidence bands. Hence these confidence sets and simultaneous confidence bands are
closely related.

It is noteworthy that the sample size $n$ only needs to satisfy $\nu = n-p-1 \ge 1$, i.e. $n \ge p+2$, so that the
regression coefficients $\bbeta$ and the error variance $\sigma^2$ can be estimated. So long as $n \ge p+2$, the theorem in Section 2
holds. A larger sample size $n$ will make the confidence sets closer to the level set, which is similar to the usual confidence sets for
the mean of a normally-distributed population. Hence the method for linear regression provided in this paper is much simpler than
that for nonparametric regression and density level sets (cf. Mammend and Polonik, 2013, Chen {\it et al.}, 2017, Qiao and Polonik, 2019).

In the theorem in Section 2, the minimum coverage probability over the whole parameter space $\bbeta \in \Re^{p+1}$ and $\sigma >0$
is sought since no assumption is made about any prior information on $\bbeta$ or $\sigma >0$. If it is known {\it a priori} that
$\bbeta$ and $\sigma$ are in a restricted space, then the usual estimators $\hat{\bbeta}$ and $\hat{\sigma}$ should be replaced by the maximum likelihood estimators over the restricted space, and the minimum coverage probability should also be over this restricted
space. This situation becomes more complicated and is beyond the scope of this paper.     

It is also pointed out that the construction method is readily applicable to other parametric regression models where the mean response depends on a linear predictor through a
monotonic link function. Examples are generalized linear models, linear mixed models
and generalized linear mixed models. The illustrative Example 3 involves a generalized linear model.
Therefore the method proposed in this paper is widely applicable.

We are unable to establish thus far whether the two-sided confidence set $[\hG_{2l}, \hG_{2u}]$ is of confidence level
$1- \alpha$ exactly. Construction of a two-sided confidence set of exact confidence level
$1- \alpha$ is clearly of interest and warrants further research. We are actively researching on this and hope to report the
results in the near future.

%It is also interesting to explore whether the construction method for linear regression in this paper, that is, using simultaneous confidence bands to construct confidence sets, could be extended to nonparametric regression. While beyond the scope of this paper, it warrants further research. 

\section{ Appendix}

In this appendix a proof of the Theorem in Section 2 is sketched. 

For proving the statement in (7), we have 
\begin{eqnarray}
&           & \left\{ \, G \subseteq \hG_{1u} \, \right\} \nonumber \\
& =         & \left\{ \, \forall \, \bx \in G: \, \bx \in \hG_{1u} \, \right\} \nonumber \\
& =         & \left\{ \, \forall \, \bx \in G: \, \ \tbx^T \hat{\bbeta} + c_1 \hat{\sigma} m( \bx ) \ge \lambda   \, \right\} \nonumber \\
& =         & \left\{ \, \forall \, \bx \in G: \, \ \tbx^T (\hat{\bbeta} - \bbeta) + c_1 \hat{\sigma} m( \bx ) \ge \lambda - \tbx^T \bbeta  \, \right\} \nonumber \\
& \supseteq & \left\{ \, \forall \, \bx \in G: \, \ \tbx^T (\hat{\bbeta} - \bbeta) + c_1 \hat{\sigma} m( \bx ) \ge 0 \, \right\} \nonumber \\
& \supseteq & \left\{ \, \forall \, \bx \in K: \, \ \tbx^T (\hat{\bbeta} - \bbeta) + c_1 \hat{\sigma} m( \bx ) \ge 0  \, \right\} \nonumber 
\end{eqnarray}
where the second equation follows directly from the definition of $\hG_{1u}$, the first ``$\supseteq$'' follows directly from the definition of $G$,
and the second ``$\supseteq$'' follows directly from the fact that $G \subseteq K$.
It follows therefore
\begin{equation}
\Pr \left\{ \, G \subseteq \hG_{1u} \, \right\} 
\ge     \Pr \left\{ \, \forall \, \bx \in K: \, \ \tbx^T (\hat{\bbeta} - \bbeta) + c_1 \hat{\sigma} m( \bx ) \ge 0  \, \right\} 
=         1-\alpha
\end{equation}
where the last equality is directly due to the fact that $\tbx^T \hat{\bbeta} + c_1 \hat{\sigma} m( \bx )$ is an upper simultaneous confidence band for 
$\tbx^T {\bbeta}$ over $\bx \in K$ of exact $1-\alpha$ level, as given in (1). 

Next we show that the minimum probability over $\bbeta \in \Re^{p+1}$ and $ \sigma >0$ in statement (7) is
$1-\alpha$, attained at $\bbeta = (\lambda, 0, \ldots 0)^T$. At $\bbeta = (\lambda, 0, \ldots 0)^T$, we have $G = K$ and so
\begin{eqnarray}
&           & \left\{ \, G \subseteq \hG_{1u} \, \right\} \nonumber \\
& =         & \left\{ \, \forall \, \bx \in K: \, \bx \in \hG_{1u} \, \right\} \nonumber \\
& =         & \left\{ \, \forall \, \bx \in K: \, \ \tbx^T \hat{\bbeta} + c_1 \hat{\sigma} m( \bx ) \ge \lambda   \, \right\} \nonumber \\
& =         & \left\{ \, \forall \, \bx \in K: \, \ \tbx^T (\hat{\bbeta} - \bbeta) + c_1 \hat{\sigma} m( \bx ) \ge \lambda - \tbx^T \bbeta  \, \right\} \nonumber \\
& =         & \left\{ \, \forall \, \bx \in K: \, \ \tbx^T (\hat{\bbeta} - \bbeta) + c_1 \hat{\sigma} m( \bx ) \ge 0  \, \right\} \, \nonumber 
\end{eqnarray}
which gives
\begin{equation}
\Pr \left\{ \, G \subseteq \hG_{1u} \, \right\} 
= \Pr \left\{ \, \forall \, \bx \in K: \, \ \tbx^T (\hat{\bbeta} - \bbeta) + c_1 \hat{\sigma} m( \bx ) \ge 0  \, \right\} = 1-\alpha \, .
\end{equation}
The combination of (11) and (12) proves the statement in (7).

Now we prove the statement in (8). For a given set $A \subseteq K$, let $A^c$ denote the complement set within $K$, i.e.
$A^c = K \backslash A$. We have 
\begin{eqnarray}
&           & \left\{ \, \hG_{1l} \subseteq G \, \right\} \nonumber \\
& =         & \left\{ \, G^c \subseteq \hG_{1l}^c \, \right\} \nonumber \\
& =         & \left\{ \, \forall \, \bx \in G^c: \, \bx \in \hG_{1l}^c \, \right\} \nonumber \\
& =         & \left\{ \, \forall \, \bx \in G^c: \, \ \tbx^T \hat{\bbeta} - c_1 \hat{\sigma} m( \bx ) < \lambda   \, \right\} \nonumber \\
& =         & \left\{ \, \forall \, \bx \in G^c: \, \ \tbx^T (\hat{\bbeta} - \bbeta) - c_1 \hat{\sigma} m( \bx ) < \lambda - \tbx^T \bbeta  \, \right\} \nonumber \\
& \supseteq & \left\{ \, \forall \, \bx \in G^c: \, \ \tbx^T (\hat{\bbeta} - \bbeta) - c_1 \hat{\sigma} m( \bx ) \le 0 \, \right\} \nonumber \\
& \supseteq & \left\{ \, \forall \, \bx \in K: \, \ \tbx^T (\hat{\bbeta} - \bbeta) - c_1 \hat{\sigma} m( \bx ) \le 0  \, \right\} \nonumber 
\end{eqnarray}
where the third equation follows directly from the definition of $\hG_{1l}$ (or $\hG_{1l}^c$), the first ``$\supseteq$'' follows directly from the definition of $G$ (or $G^c$),
and the second ``$\supseteq$'' follows directly from the fact that $G^c \subseteq K$.
It follows therefore
\begin{equation}
\Pr \left\{ \, \hG_{1l} \subseteq G \, \right\} 
\ge     \Pr \left\{ \, \forall \, \bx \in K: \, \ \tbx^T (\hat{\bbeta} - \bbeta) - c_1 \hat{\sigma} m( \bx ) \le 0  \, \right\} 
=         1-\alpha
\end{equation}
where the last equality is directly due to the fact that $\tbx^T \hat{\bbeta} - c_1 \hat{\sigma} m( \bx )$ is a lower simultaneous confidence band for 
$\tbx^T {\bbeta}$ over $\bx \in K$ of exact $1-\alpha$ level, as given in (2). 

Next we show that the minimum probability over $\bbeta \in \Re^{p+1}$ and $ \sigma >0$ in statement (8) is
$1-\alpha$, attained at $\bbeta = (\lambda^-, 0, \ldots 0)^T$, where $\lambda^-$ denotes
a number that is infinitesimally smaller than $\lambda$. At $\bbeta = (\lambda^-, 0, \ldots 0)^T$, we have $G^c = K$ and so
\begin{eqnarray}
&           & \left\{ \, \hG_{1l} \subseteq G \, \right\} \nonumber \\
&           & \left\{ \, G^c \subseteq \hG_{1l}^c \, \right\} \nonumber \\
& =         & \left\{ \, \forall \, \bx \in G^c: \, \bx \in \hG_{1l}^c \, \right\} \nonumber \\
& =         & \left\{ \, \forall \, \bx \in K: \,   \bx \in \hG_{1l}^c \, \right\} \nonumber \\
& =         & \left\{ \, \forall \, \bx \in K: \, \ \tbx^T \hat{\bbeta} - c_1 \hat{\sigma} m( \bx ) < \lambda   \, \right\} \nonumber \\
& =         & \left\{ \, \forall \, \bx \in K: \, \ \tbx^T (\hat{\bbeta} - \bbeta) - c_1 \hat{\sigma} m( \bx ) < \lambda - \tbx^T \bbeta  \, \right\} \nonumber \\
& =         & \left\{ \, \forall \, \bx \in K: \, \ \tbx^T (\hat{\bbeta} - \bbeta) - c_1 \hat{\sigma} m( \bx ) < 0  \, \right\} \, \nonumber 
\end{eqnarray}
which gives
\begin{equation}
\Pr \left\{ \, \hG_{1l} \subseteq G \, \right\} 
= \Pr \left\{ \, \forall \, \bx \in K: \, \ \tbx^T (\hat{\bbeta} - \bbeta) - c_1 \hat{\sigma} m( \bx ) < 0  \, \right\} = 1-\alpha \, .
\end{equation}
The combination of (13) and (14) proves the statement in (8).

The statement (9) can be proved by combining the arguments that establish (11) and (13) above to establish that
$$
\left\{ \, \hG_{2l} \subseteq G \subseteq \hG_{2u} \, \right\}
\supseteq
\left\{ \, \forall \, \bx \in K: \, \ -c_2 \hat{\sigma} m( \bx ) \le \tbx^T (\hat{\bbeta} - \bbeta) < c_2 \hat{\sigma} m( \bx )   \, \right\};
$$
details are omitted here to save space. Unfortunately, a least favorable configuration of $\bbeta$ that achieves the coverage probability $1-\alpha$ cannot be identified 
in this case, and so $1-\alpha$ is only a lower bound on the
confidence level.

\bibliographystyle{rss}
\renewcommand{\bibname}{References}

\label{lastpage}

\end{document}